\documentclass[3p,times]{elsarticle}

\usepackage{amsmath}
\usepackage{amssymb}
\usepackage{accents}
\usepackage{bm}
\usepackage{graphicx}
\usepackage{color}
\usepackage{dsfont}
\usepackage{array}
\usepackage{geometry}
\usepackage{amsfonts}
\usepackage{longtable}
\usepackage[table]{xcolor}
\usepackage{amssymb}
\usepackage[figuresright]{rotating}

\def\MPlh{M_{_{\rm Pl}}}
\begin{document}
\begin{frontmatter}



\title{Low Scale Higgs Inflation with Gauss-Bonnet Coupling}


\author{Jose Mathew and S. Shankaranarayanan}

\address{Indian Institute of Science Education and Research, Thiruvananthapuram}

\begin{abstract}
  Recent LHC data provides precise values of coupling constants of the
  Higgs field, however, these measurements do not determine its
  coupling with gravity.  We explore this freedom to see whether Higgs
  field non-minimally coupled to Gauss-Bonnet term in 4-dimensions can
  lead to inflation generating the observed density fluctuations.  We
  obtain analytical solution for this model and that the exit of
  inflation (with a finite number of e-folding) demands that the
  energy scale of inflation is close to Electro-weak scale. We compare
  the scalar and tensor power spectrum of our model with PLANCK data
  and discuss its implications.
\end{abstract}


\end{frontmatter}


\section{Introduction}
Cosmological Inflation ~\cite{book:15483,book:15485,book:15486,book:75690,mukhanov1992theory}
provides a causal mechanism to generate the primordial density
perturbations that are responsible for the anisotropies in the cosmic
microwave background~(CMB) and the formation of the large scale
structure~(LSS).  CMB and LSS data have been used to constrain the
parameters of the inflationary model. In the case of canonical scalar
field, the CMB and LSS data provide constraints on the height and the
slope of the potential~ref~\cite{lidsey1997reconstructing,bassett2006inflation,ade2014planck}.

The fact that the temperature fluctuations of the CMB is close to
scale-invariance is a highly demanding requirement for inflation model
building ~ref~\cite{lyth1999particle,lyth2008particle,mazumdar2010particle,yamaguchi2011supergravity} than
providing approximately $60$ e-foldings of inflation needed to solve
the various initial conditions problems.  More specifically, the near
scale-invariance imposes a condition that the canonical scalar field
potential should be almost flat --- almost like cosmological constant
--- so that the quantum fluctuations that exit the horizon during
inflation is nearly scale-invariant. While these flat potentials are
phenomenologically successful, however, in the standard model of particle physics there is no candidate with such flat potentials that could sustain inflation~\cite{lyth1999particle,lyth2008particle,mazumdar2010particle,yamaguchi2011supergravity}.  For instance, the
renormalizability of the Higgs field in 4-dimensions puts a constraint
on the scalar field potential ($V(\phi) = m^2 \phi^2 + \lambda
\phi^4$, where $m$ is the mass and $\lambda$ is the coupling
parameter), however, inflationary models require potentials of the
form $V(\phi) = \sum_{n = 0}^{N} c_{2n} \phi^{2n}$ where $c_{2n}$'s
are real numbers and $N > 2$.

To achieve the flatness of the potential, inflationary models using the standard model Higgs field as the inflaton, couples Higgs field non-minimally with
gravity~\cite{cervantes1995induced1,cervantes1995induced2,bezrukov2008standard,germani2010new,cai2015higgs,van2016higgs}.
In Higgs Inflation 
~\cite{cervantes1995induced1,cervantes1995induced2,bezrukov2008standard,barvinsky2008inflation,bezrukov2009standard,barvinsky2009asymptotic,de2009running,bezrukov2011higgs,barvinsky2012higgs,salvio2013higgs},  the flatness of the Higgs potential is achieved by large non-minimal coupling of the Higgs field to the Ricci scalar ($\sim \xi R \phi^2$ where $\xi$ is the coupling constant and $R$ is the Ricci scalar) i. e. $\xi \sim 10^4$.

One of the main assumption in the above models of Higgs inflation is
that the standard model physics remains to hold until Planck
energy. Which may be consistent with the current LHC
measurements --- since there are no evidence of new physics so far
(e.g., supersymmetry or extra dimension(s), etc.) 
~ref~\cite{aad2012observation,chatrchyan2013observation,aad2015combined} ---
however, it also points to the fact that $\lambda$ can be negative at high energies \cite{bezrukov2012higgs,degrassi2012higgs,buttazzo2013investigating,branchina2013stability,branchina2014top,espinosa2015implications,espinosa2015cosmological}. But a non-minimal Higgs Ricci coupling may prevent this up to inflationary scale \cite{bezrukov2009standard,bezrukov2011higgs}.

In this work, we ask the following question: Can Higgs field drive
inflation without invoking any new Physics in the particle physics
sector with exit at low-energies, order of 100 GeV to 1000 GeV? While
the LHC measurements determine the coupling constants of the Higgs
field precisely, it does not determine its coupling with Gravity. We
use this freedom and assume that the Higgs field couples with the
Gauss-Bonnet Gravity, instead of Ricci Scalar.

Gauss-Bonnet Gravity is a part of the general extension of Gravity
theories referred to as Lovelock theories of gravity
~\cite{lovelock1971einstein}. One important feature of Lovelock theories, as
against the $f(R)$ gravity theories, is that the gravity equations of
motion remain second order (and quasilinear in second order). They
provide a natural arena for understanding many deep features of
gravity and recently they have been a subject of study. (For a recent review see ~\cite{padmanabhan2013lanczos}.) Some higher dimensional Lovelock theories arise also as a weak field limit of string theory
\cite{gross1986superstring,gross1987quartic}. While a pure Gauss-Bonnet term is
non-dynamical in 4-dimensions --- topologically invariant in
4-dimensions --- non-minimal coupling with the Higgs field makes it
dynamical.

Since Gauss-Bonnet term is higher-derivative term, it may be natural
to expect that non-minimal coupling of the scalar field may only lead
to modifications at high-energies. However, in this work, we show
explicitly that the non-minimal coupling of the Higgs scalar lead to
exit of inflation at low-energies i.e. close to Electroweak
scale. This an unique feature of our model. We also explicitly compute
the power-spectrum and show that it is consistent with the recent
PLANCK data \cite{ade2014planck}. There has been recent interest in
coupling scalar field with Gauss-Bonnet gravity (see, for instance,
\cite{kanti2015gauss,van2016higgs}). Our analysis differs from their
analyses: In Ref. \cite{kanti2015gauss}, authors have assumed that
Einstein-Hilbert term is irrelevant and, hence, have ignored the
linear term. In Ref. \cite{van2016higgs}, the authors have coupled
the scalar field to both the Ricci and Gauss-Bonnet gravity. Their analysis is 
based on slow-roll and makes predictions similar to \cite{bezrukov2008standard}. 
It is important to note that they found the Gauss-Bonnet term to be significant only 
at late times where as the Higgs-Ricci coupling dominating the initial epoch and was responsible for the spectrum. 
As mentioned earlier, our model couples the Higgs scalar with Gauss-Bonnet gravity leading to a dynamical model of inflation.

The paper is organized as follows: In the next section, to get the
physical picture of the dynamical equations, we obtain exact
generalized power-law inflation for our model. We show that the
generalized power-law solution exists only when the mass of the scalar
field is identically zero. In Sec. \ref{sec:Higgsinflation}, we show
that the Higgs potential leads to dynamical model of inflation where
the exit occurs close to the electro-weak scale. We show that the
non-zero Higgs mass lead to the exit. In Sec. \ref{sec:Pspectrum}, we
compute the power-spectrum of our model and compare the results with
the recent PLANCK data. We discuss the key results and possible
implications of our model in Sec. \ref{sec:Conclusions}.

In this work, we consider $(-, +, + ,+)$ metric signature. We use
natural units $c=\hbar=1$, $\kappa=1/\MPlh^2$, and $\MPlh^2= \frac{\hbar
c}{8 \pi G}$ is the reduced Planck mass. We denote dot as derivative with
respect to cosmic time $t$ and $H(t) \equiv \dot{a}(t)/a(t)$.
%We denote ${}^\prime$ as derivative with respect to scalar field ($\phi$).

%------------------------------------------------
\section{Generalized power-law inflation}
\label{sec:PLInflation}

Consider the following action where the scalar field $\phi$ is
non-minimally coupled of Gauss-Bonnet ($ \mathcal{L}_{\rm GB}$) term:
\begin{eqnarray}  
{\cal S}=\int d^4x \sqrt{-g} \left[\frac{R}{2\kappa} +
f(\phi)\mathcal{L}_{\rm GB}-\frac{1}{2} \nabla_{a} \phi \nabla^{a} \phi -V\left(\phi\right)\right],%+{\cal S}_m
\label{eq:action}
\end{eqnarray} 
where $R$ is the Ricci scalar, $V(\phi)$ is the scalar field
potential, $f(\phi)$ is the coupling parameter and
\begin{equation}
 \mathcal{L}_{GB} = R^2-4R^{ab}R_{ab} + R^{abcd}\, R_{abcd} 
\end{equation}
Varying the action (\ref{eq:action}) w.r.t the field $\phi$ and the
metric leads to the following equations of motion \cite{nojiri2005gauss}:
\begin{equation}
\Box \phi +  \mathcal{L}_{GB} \, f_{,\phi} \left( \phi \right) - V_{,\phi} \left( \phi \right)=0
\label{eq:PhiEOM}
\end{equation}
%
%%%%%%%%%%%%
\begin{eqnarray}
 && \frac{1}{\kappa} G_{p q}=\left(8G_{p q}g^{a b}\nabla_{a b}{f\left(
        \phi \right)}+4R\nabla_{pq}{f\left(\phi\right)} -
    8R^{a}_{p}\nabla_{aq}{f\left(\phi\right)} -8R^a_{q} \nabla_{a
      p}{f\left( \phi \right)} \right.\nonumber \\ 
      && \left.  8\nabla_{a b}{f\left( \phi
      \right)}R^{a b}g_{p q} 
      -8\nabla_{a b}{f\left(\phi\right)}R_{p \;
      q}^{\;\; a \; b}+ \nabla_{p}{\phi}\nabla_{q}{\phi} - g_{p
      q}\left(\frac{1}{2}g^{a b}
      \nabla_{a}{\phi}\nabla_{b}{\phi}+V\left(\phi\right)\right)
  \right)\, 
\label{eq:EinsEOM}
\end{eqnarray}
%%%%%%%%%%%%%%%%
It must be noted that the field equations being second order implies
that this model doesn't have the problem of unitarity.

In this section, we are interested in obtaining exact solution for the
above set of equations of motion for a spatially flat
Friedmann-Robertson-Walker (FRW) background
\begin{equation}
ds^2=-dt^2 + a(t)^2 \left( dx^2 + dy^2 + dz^2 \right)
\end{equation}
where $a(t)$ is the scale factor.  The equation of the field $\phi(t)$
and the scale factor $a(t)$ follows from
Eqs. (\ref{eq:PhiEOM},\ref{eq:EinsEOM}), respectively
%%%%%%%%%%
\begin{subequations}
\label{eq:threeequations}
\begin{eqnarray}
\label{eom}
 - 24 H^2 \frac{\ddot{a}}{a} {f}_{,\phi} \left(\phi\right) +
  \ddot{\phi}+V_{,\phi}\left(\phi\right)+ 3 H \dot{\phi} &=& 0 \\ 
%%%
\label{energyconstraint}
- 3 H^2 \left(\frac{1}{\kappa} +   8 H \dot{f}\left(\phi\right) \right) 
+ \frac{1}{2}\dot{\phi}^2 + V\left(\phi\right) &=& 0 \\
%%%
\label{evolutionequation}
-H^2 \left( \frac{1}{\kappa} + 8 \ddot{f}\left(\phi\right) \right)
- \frac{2 \ddot{a}}{a} \left( \frac{1}{\kappa} +  8 H \dot{f}\left(\phi\right) \right) 
+V\left(\phi\right) -\frac{1}{2} \dot{\phi}^2  &=& 0
\end{eqnarray}
\end{subequations}
It is important to note that the above differential equations are
quasilinear i.e.  they are linear with respect to all the highest
order derivatives of $a(t)$ and $\phi(t)$.  Rewriting Eqs.~(\ref{eom},
\ref{energyconstraint}), we get
\begin{eqnarray}
&& -2 H^2 + \kappa\dot{\phi}^2 - 24 \, \kappa \, H^3\dot{f}\left(\phi\right) + 2 \frac{\ddot{a}}{a} + 
16 \, \kappa H \frac{\ddot{a}}{a} \dot{f}\left(\phi\right) \nonumber \\
&&+ 8\kappa H^2\ddot{f}\left(\phi\right)=0
\label{eq:keyeqn}
\end{eqnarray}
In the rest of this section, we consider the solution of
(\ref{eq:threeequations}) for the following ansatz 
\begin{equation}
\label{eq:PLansatz}
a(t) =a_0\left(\frac{t}{t_0} + \Upsilon \right)^p~~\mbox{and}~~ \phi=\phi_0 \left(\frac{t}{t_0} + \Upsilon \right)^n
\end{equation}
where $p > 1$ is a constant; $n$ is a constant;  $a_0$, $t_0$ are arbitrary 
constants whose values will not appear in any physical measured quantities and $\Upsilon$ 
is given by 

\begin{equation}
\Upsilon = \left(\frac{\phi(t_0)}{\phi_0}\right)^{1/n} - 1 \, .
\end{equation}
Usually in cosmology, 
power-law inflation is given by $a(t) \propto t^p$. Ansatz (8) is a generalization. For 
real integer $p$, we have 
$$a(t) = a_p t^p + a_{p - 1} \, t^{p -1} + a_{p - 2} \, t^{p -2} + \cdots + a_0$$ 
where in our case the coefficients $a_p, a_{p - 1},\cdots a_0$ are related. Since, $a(t)$ is a 
series, $\phi$ should also be a series like 
$$ \phi(t)  = \phi_n t^n + \phi_{n -1}  \, t^{n -1} + \phi_{n - 2} \, t^{n -2} + \cdots + \phi_0$$ 
where, again, all the coefficients  $\phi_n, \phi_{n -1},\cdots \phi_0$. At $t = t_0$, 
$ \phi(t_0) \neq \phi_0$ and $\phi_0$ is an independent parameter. We refer to the above ansatz (\ref{eq:PLansatz}) for the scale factor as generalized power-law inflation.

Substituting the above ansatz (\ref{eq:PLansatz}) in
Eq.~(\ref{eq:threeequations}), we get the following {\sl exact
  relations}
\begin{subequations}
\label{eq:PLgensol}
\begin{eqnarray}
  \label{eq:PLgensol1}
V\left(\phi\right) &=& \tilde{\lambda}_1 \; {\phi}^{-\frac{2}{n}} 
\!\!\! \; \; + \; \tilde{\lambda}_2 \; {\phi}^{\frac{2(n-1)}{n}}
\!\!\! \; \; \; + \; \tilde{\lambda}_3 \; {\phi}^{\frac{p-1}{n}}   \\ 
\label{eq:PLgensol2}
f\left(\phi\right) &=& \; \tilde{\alpha}_1 \; {\phi}^{\frac{2}{n}} 
\; + \; \tilde{\alpha}_2 \; {\phi}^{\frac{2(n+1)}{n}} 
\; + \; \tilde{\alpha}_3 \; {\phi}^{\frac{3+p}{n}}
\end{eqnarray}
\end{subequations}
where
\begin{equation}
\label{eq:PLgencoefficients}
\begin{split}
\tilde{\lambda}_1 &= \frac{3\left(p -1\right)p^2} {\kappa \left(p+1 \right)} \left(\frac{\phi_0^{1/n}}{t_0}\right)^2\quad \quad \quad 
\tilde{\lambda}_2  =  \frac{(5n^2p-n^2+2n^3)}{2(1-2n+p)} \left(\frac{\phi_0^{1/n}}{t_0}\right)^2 \quad \quad \quad \quad \quad\quad
\tilde{\lambda}_3  = 24 p^3C \left(\frac{\phi_0^{1/n}}{t_0}\right)^{1-p}
\\
\tilde{\alpha}_1  &=   \frac{-1}{8\kappa p(1+p)}  \left(\frac{\phi_0^{1/n}}{t_0}\right)^{-2} \quad \quad \quad 
\tilde{\alpha}_2   = \frac{n^2 }{16p^2(1+n)(1-2n+p)}\left(\frac{\phi_0^{1/n}}{t_0}\right)^{-2} \quad \quad \quad 
\tilde{\alpha}_3   =   \frac{C}{p+3}\left(\frac{\phi_0^{1/n}}{t_0}\right)^{-(p+3)}
\end{split}    
\end{equation}
and $C$ is the constant of integration. 
\\ \\

The following points are important to note regarding the above generalized power-law solution: 
(i) The  ansatz (\ref{eq:PLansatz}) is the most general power-law {\sl exact} solution satisfying Eqs.~(\ref{eq:threeequations}) for the potential and coupling (\ref{eq:PLgensol}) and {\it does not} depend on
the constant of integration $C$.
(ii)  The above solutions are valid for any $p > 1$ and $n$. Imposing the condition that the 
potential be non-negative leads to $n > -2$ and $C \geq 0$. 
(iii) The coefficient of the first term in RHS of (\ref{eq:PLgensol1}) dominates the coefficients 
of the other two terms. Similarly, the coefficient of the first term in RHS of (\ref{eq:PLgensol2}) 
dominates the coefficients of the other two terms:

\begin{equation}
\label{dominance}
\left|\frac{\tilde{\alpha}_2}{\tilde{\alpha}_1}\right|=\left|\frac{1}{\MPlh^2}\frac{ n^2 (p+1)}{2 (n+1) p^2 (-2 n+p+1)}\right| \quad \quad \quad
\left|\frac{\tilde{\lambda_2}}{\tilde{\lambda_1}}\right|=\left|\frac{1}{\MPlh^2} \frac{ n^2 (p+1) (2 n+5 p-1)}{6 (p-1) p^2 (-2 n+p+1)}\right|
\end{equation}
Hence for $\phi \ll \MPlh$;~~~ $\tilde{\lambda}_1 \phi^{-2/n}$ and $\tilde{\alpha}_1 \phi^{2/n}$ are the dominant terms in the potential and coupling respectively
(iv)  In the case of $C =0$ and $\phi \ll \MPlh$, the  
general solution (\ref{eq:PLgensol}) leads to  
$$V(\phi)=\tilde{\lambda}_1\phi^{\tilde{n}} \qquad f(\phi)=\tilde{\alpha}_1 \phi^{-\tilde{n}} \qquad \mbox{where}~~\tilde{n} = -2/n $$
and is identical to the one studied by Guo and Schwarz \cite{guo2010slow,jiang2013inflation}. 
The authors~\cite{guo2010slow} claimed that the solution has a graceful exit. 
However,  our analysis clearly indicate that the above solution is a subset of the 
generalized power-law inflationary solution where there is no exit from inflation.

To make it clear, let us now repeat the analysis of  \cite{guo2010slow} in the limit $\phi \ll \MPlh$. 
Rewriting their variables in terms of our variables, we have 
~~$\omega = 1$,  ~~$V_0 =\tilde{\lambda}_1$,~~  $\xi_0=-2 \tilde{\alpha}_1$,    
~~$\alpha=-(8/3) \tilde{\alpha}_1 \tilde{\lambda}_1$.  Hence from Eq.~(\ref{eq:PLgencoefficients}), ~ we get ~$\alpha =
\frac{p(p-1)}{(p+1)^2}$. Now ~~$0<\alpha<1$ implies $p>1$, i.e at late times when our approximations turn valid 
the solution asymptotically approaches the power-law solution (\ref{eq:PLansatz}). For $\alpha=0.25$ our analysis 
shows that the slow-roll parameter asymptotically takes the value $\epsilon=0.4641$, which is also obtained numerically
as the late time behaviour by Bruck and Longden  \cite{van2016higgs} [see Fig.~(1) in that reference].

In the rest of this section, we consider the special case $C = 0$, however,
without imposing the condition $\phi \ll \MPlh$ and find an
interesting scenario for Higgs inflation.

%------------------------------------------------
\subsection{Special case: C = 0}
\label{sec:specialcase}

Condition on n and C are that $n > -2$ and $C \geq 0$. We are interested in looking at the case where the 
scalar field potential has the form of power-law and consistent with standard 
model of particle physics. Hence, for the special case $C = 0$ and $n = - 1/2$, scalar field potential
and the coupling function take the following simple form:
\begin{subequations}
\label{eq:PLsplsol}
\begin{eqnarray}
V\left(\phi\right)&=& \lambda_4  {\phi}^{4}  + \lambda_6 \phi^6 \\
 %%%%%%%%%%%%%%%%%%%%     
 f\left(\phi\right)&=& \frac{\alpha_2}{\phi^2} +  \frac{\alpha_4}{\phi^4} 
\end{eqnarray}
\end{subequations}
where 
\begin{equation}
\label{eq:PLcoefficients}
\lambda_4  = {\frac {3{p}^{2} \left(p-1 \right) }{{{ \phi_0}}^{4}\kappa
    \left( p+1 \right) {{t_0}}^{2}}} \quad
\lambda_6 = \frac{1}{4}\,{\frac { \left( 5\,p-2\right)}{{{ \phi_0}}^{4}{{ t_0}}^{2} \left( 4+2\,p
    \right) }} \quad 
\alpha_2 =      \frac{1}{32}\,{\frac{{{\phi_0}}^{4}{{ t_0}}^{2}}{{p}^{2}
    \left( 2+p \right) }}  \quad
\alpha_4 = - \frac{1}{8}\,{\frac {{{\phi_0}}^{4}{{t_0}}^{2}}{\kappa p \left( p+1 \right) }}       
\end{equation}
This is one of the main results of this work regarding which we would like to stress the following 
points: (i) The dominant term in the potential and coupling function are  $\lambda_4$ and $\alpha_4$, respectively: 

\begin{equation}
\left|\frac{\lambda_4}{\lambda_6}\right| \approx \left|5p^2 \MPlh^2\right| \quad \quad \quad \quad \quad \quad%
\left|\frac{\alpha_4}{\alpha_2}\right|\approx \left|4 p \MPlh^2\right| \, .
\label{dominance2}
\end{equation}
 In other words, approximating
\begin{equation}
\label{eq:PLapprox}
V(\phi) \simeq \lambda_4  {\phi}^{4} \qquad f\left(\phi\right) \simeq \frac{\alpha_4}{\phi^4}
\end{equation}
lead to power-law expansion. {While (\ref{eq:PLansatz}) is an exact solution of the background field equations 
(\ref{eq:threeequations}) for the form of potential and coupling given by (\ref{eq:PLsplsol}), including the sub-dominant terms. 
The ansatz (\ref{eq:PLansatz}) is an approximate solution for the form of potential and coupling given by (\ref{eq:PLapprox}).} 
It is interesting to note that the above approximate scalar field potential is the potential for the
chaotic inflation~\cite{linde1983chaotic}.  In the case of chaotic
inflation, the scalar field is not coupled to the Gauss-Bonnet
gravity, here, the non-minimal coupling catalyzes scalar field to
accelerate and, hence, there is no exit with a pure $\phi^4$
potential.
(ii) For inflation to occur, Eq.~(\ref{eq:PLcoefficients}) implies
$\alpha_4 < 0$ and $\lambda_4 > 0$.
(iii) Physical relevance of $\phi_0$ can be seen from
Eq. (\ref{eq:PLcoefficients}).  The value of Gauss-Bonnet coupling
parameter at the epoch of inflation depends on the ratio of $\phi_0$
and $\phi(t_0)$, i. e.,
\begin{equation}
  f(\phi(t_0)) \propto \left(\frac{\phi_0}{\phi(t_0)}\right)^4 \frac {1}{\kappa} 
\end{equation}
$\phi_0$ decides the epoch of inflation in our model. Since, this can
not be fixed {\it ab initio}, it can be fixed from observations.
(iv) From (\ref{eq:PLcoefficients}), we obtain a relation between
$\lambda_4$ and $\alpha_4$
\begin{equation}
\lambda_4 |\alpha_4| = \frac{3p(p-1)}{8(p+1)^2\kappa^2}
\label{alpha4lambda4relation}
\end{equation}
The above relation shows that for a given power-law inflation,
$\lambda_4$ and $\alpha_4$ are inversely related to each other.  We
show that the above relation is approximately satisfied also for Higgs
inflation.  This is interesting because $\lambda_4$ is measured
precisely, and our model makes a precise prediction for $\alpha_4$ and
can lead to potential signatures at the LHC. We discuss this in the
Conclusion.
(v) The above relations also provide interesting connection between
$\phi_0$ (the initial value of the field) with the coupling constants.
From (\ref{eq:PLcoefficients}), we get,
\begin{equation}
\phi_0^2  = \frac{1}{t_0} \left( 24 p^3 (p - 1) \frac{|\alpha_4|}{\lambda_4}  \right)^{1/4} 
\end{equation}
Our model makes a precise prediction on the initial value of the
inflaton field once we fix the inflation epoch of inflation and hence,
setting the scale of inflation. Our model does not impose any
restriction on the epoch of power-law inflation. As we show explicitly
in the next section, including the mass term in the potential leads to
exit from inflation, interestingly, the exit occurs at low-energies.

\section{Higgs inflation}
\label{sec:Higgsinflation}

Following the discussion in the previous section, the tree-level SM
Higgs Lagrangian is
\begin{equation}
\label{eq:Higgsaction}
S =\int {\rm d}^4 x \sqrt{-g} \left[\frac{R}{2 \kappa}  + f({\cal H}) {\cal L}_{\rm GB} - |D_\mu {\cal H}|^2 - \lambda (|{\cal H}|^2-v^2)^2 \right], 
\end{equation}
where $D_\mu$ is the covariant derivative with respect to the SM gauge
symmetry, ${\cal H}$ is the SM Higgs boson, $v$ is its vacuum
expectation value ($v=246 \;\mbox{GeV}$), and $\lambda$ is the self-coupling constant.

Taking the gauge
$^t{\cal H} =(0,v+ h)/\sqrt{2}$ (where $h$ stands for the real,
neutral component of the Higgs doublet ${\cal H}$) --- the action
(\ref{eq:Higgsaction}) becomes
\begin{equation}
\label{eq:actionHiggs}
S = \int d^4 x \sqrt{-g} \left[\frac{R}{2\kappa} + f(h) {\cal L}_{\rm GB} - \frac{1}{2} g^{\mu\nu} \partial_\mu h \partial_\nu h 
- \lambda_4 \left(h^2 - v^2\right)^2 \right], 
\end{equation}
\hspace{11 cm} where $\lambda_4=\lambda/4$
\\ \\
In the limit $h \gg v$, this is identical to the action (\ref{eq:action}) --- with the
coupling constant and potential given by (\ref{eq:PLapprox}) --- that lead to
power-law inflation.  While the condition $h \gg v$ may be valid at
the initial epoch of inflation, it may not be a good approximation at
the end of inflation. 
As in the previous section, we set the Gauss-Bonnet coupling function
to be
\begin{equation}
 f(h) = \frac{\alpha_4}{h^{4}}
\end{equation}
Before we proceed to obtain the background solutions, we would like to
point the following: (i) As mentioned above, at the initial epoch of
inflation, the value of the Higgs $h$ is much larger than the vacuum
expectation value. During this epoch, the background dynamics can be
expressed as described in the previous section.  (ii) However, as the
Universe cools down during inflation, the value of the Higgs field is
of the order of the vacuum expectation value and hence, the potential
term will be sub-dominant. This suggests that the background dynamics
may change dramatically.

Unlike the earlier case, the background equations of motion
(\ref{eq:threeequations}) can not be computed analytically especially
in the region when the field value $h$ is of the order of vacuum
expectation value $v$. We solved the equations of motion
(\ref{eq:threeequations}) numerically for a time step of $10^{-16} \; \mbox{secs}$ and
precision of the field  $h/h_0$ (in dimensionless units) to be $10^{-16}$. Using LHC results of $\lambda = 0.1291$~~($\lambda=\frac{M_H^2}{2 v^2}$, where $M_H$ is the Higgs mass and $v$ its vacuum expectation value), we have $\lambda_4=\lambda/4=0.032275$. (Appendix contains the details of the
numerics.)  

In Figures (\ref{fig:epsilonvsN} and \ref{fig:phivsN1}), we have plotted different physically relevant background quantities,
from which we infer the following:

%%%%%%%%%%%%%%
\begin{figure}[!htb]
\centering % \begin{center}/\end{center} takes some additional vertical space
\includegraphics[width=.33\textwidth,origin=c,angle=0,natwidth=610,natheight=642]{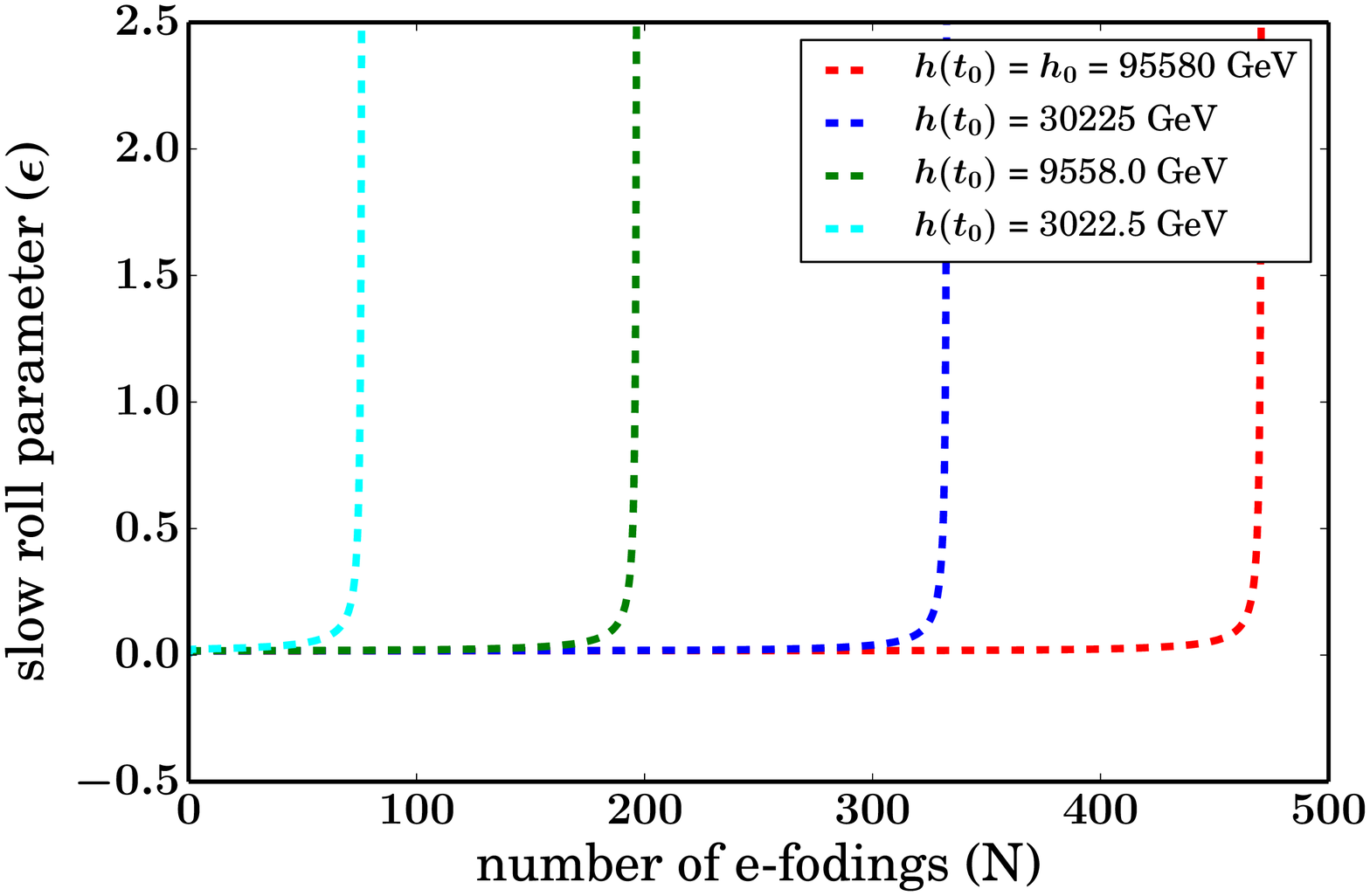}
\hfill
\includegraphics[width=.33\textwidth,origin=c,angle=0,natwidth=610,natheight=642]{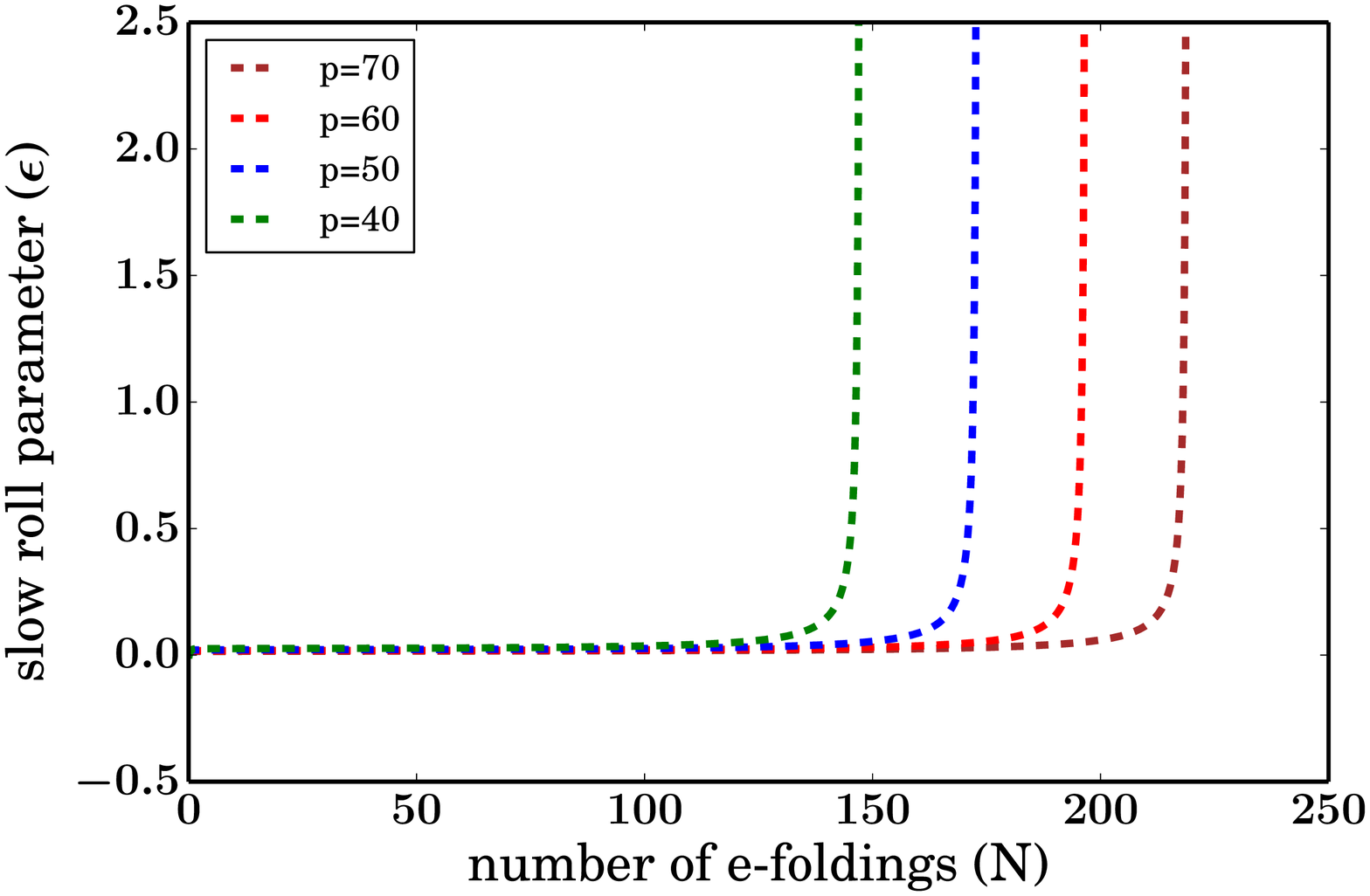}
\hfill
\includegraphics[width=.33\textwidth,origin=c,angle=0,natwidth=610,natheight=642]{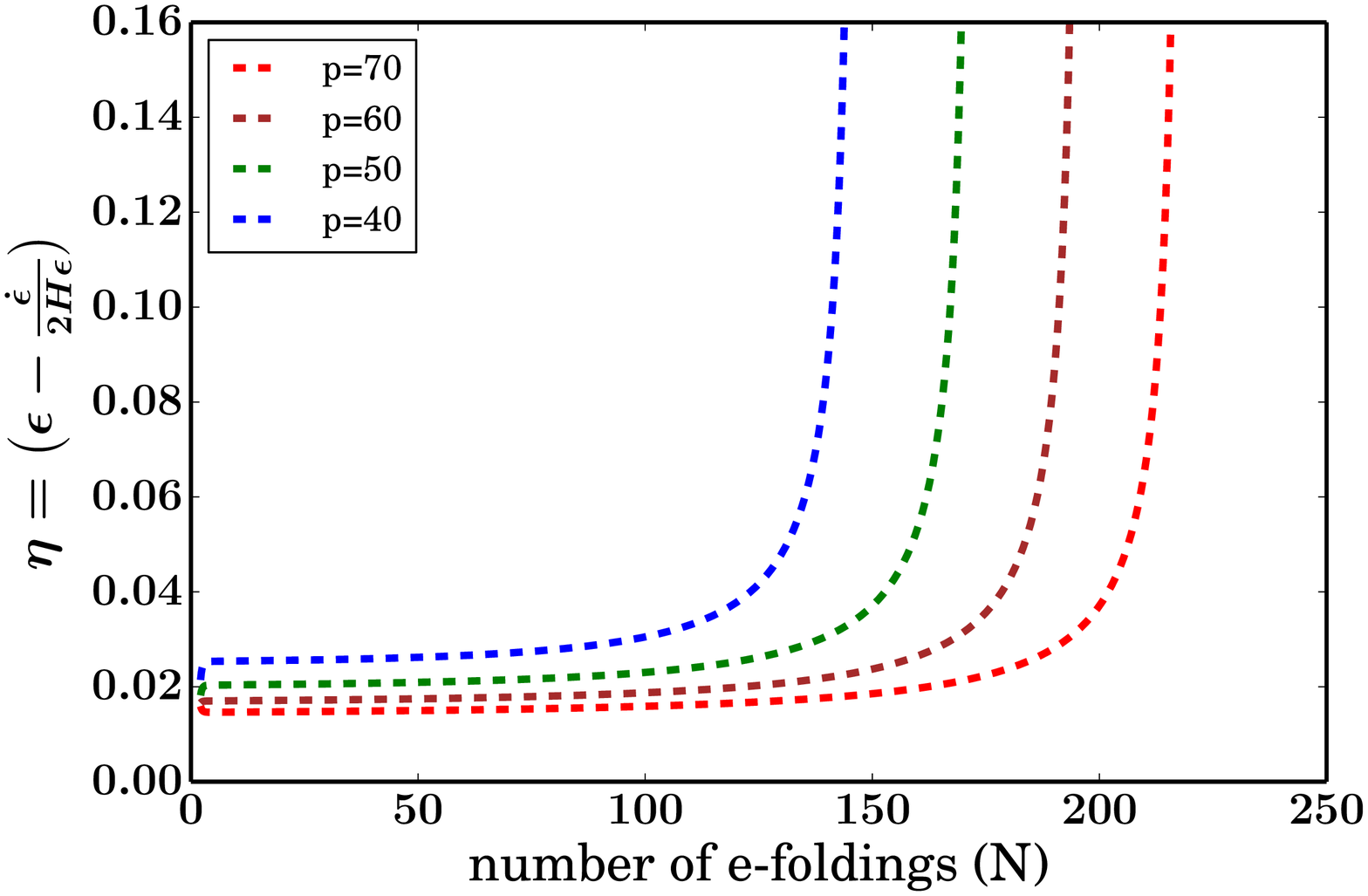}
% "\includegraphics" is very powerful; the graphicx package is already loaded
\caption{ Slow roll parameters ${\epsilon}$ and $\eta$ Vs number of e-foldings (i) $\epsilon$ for different values of Higgs field $h(t_0)$ (for p=60),(ii) $\epsilon$ for different values of $\alpha_4(p)$ and (iii) slow-roll parameter $\eta$ for different $\alpha_4(p)$}
\label{fig:epsilonvsN}
\end{figure}
\vspace{2 cm}
\begin{enumerate}
\item In Fig.~(\ref{fig:epsilonvsN}), we have plotted the slow-roll
  parameters $\epsilon$ and $\eta$
  \begin{equation}
    \epsilon = - \frac{\dot{H}}{H^2} \quad \quad \quad \quad \quad
    \eta = \epsilon - \frac{1}{2} \frac{\dot{\epsilon}}{H \epsilon}
\end{equation}
as a function of Number of e-foldings $N(t) = \ln[a(t_{end})/a(t)] =
\int_{t}^{t_{end}} H dt$ for (i) for different initial values of the field
$h(t_0)$, (ii) assuming that initial epoch as a power-law $p >
1$ and (iii) assuming that initial epoch as a power-law $p >
  1$.  This clearly indicates that our model leads to a minimum of 100 e-foldings of inflation for any initial field value $h(t_0)$, greater than $ 5~\mbox{TeV}$. Fig.~(\ref{fig:epsilonvsN}) also shows that
larger the initial value of the field it leads to longer number of
e-foldings and $\eta$ is a constant for all through the inflation and
varies rapidly at the exit of inflation.  
%
%\item In Fig.~(\ref{fig:phivsN1}), we have plotted the time evolution
% of the Higgs $h(t)$ as a function of time for different initial
%values of the field. Plots clearly show that different initial
%  values of the field converges to a same value at late times and
% shows that the model is an attractor.
%
%\item In Fig.~(\ref{fig:phivsN2}), we have plotted the time evolution
%  of the Higgs $h(t)$ as a function of time for by assuming different
%  power-law value at the initial epoch. This also show that power-law
%  behaviour assumption at the initial epoch lead to attractor.
 
 \item In Fig.~(\ref{fig:phivsN1}), we have plotted the evolution of the Higgs field $h(t)$ as a function of number of e-foldings for by assuming (i) different initial values of the field
 $h(t_0)$ and (ii) assuming that initial epoch as a power-law $p >
 1$. It is important to note, in
 our model like any other model of inflation the field value remains
 almost a constant throughout the evolution, however close to the exit
 it changes rapidly. From Figs.~(\ref{fig:phivsN1}),
  we infer that whatever be the initial value of the field or power-law
  $p$, the exit of inflation --- when $\epsilon$ becomes greater than
  $1$ --- occurs around the Electroweak scale.
\end{enumerate}
We have explicitly shown that the Higgs neutral scalar field,
non-minimally coupled to Gauss-Bonnet gravity leads to a dynamical
model of inflation and exit of Inflation occurs around Electroweak
scale $\sim 250 {\rm GeV}$.

\begin{figure}[!htb]
  \centering % \begin{center}/\end{center} takes some additional
             % vertical space
\includegraphics[width=.48\textwidth,origin=c,angle=0,natwidth=610,natheight=642]{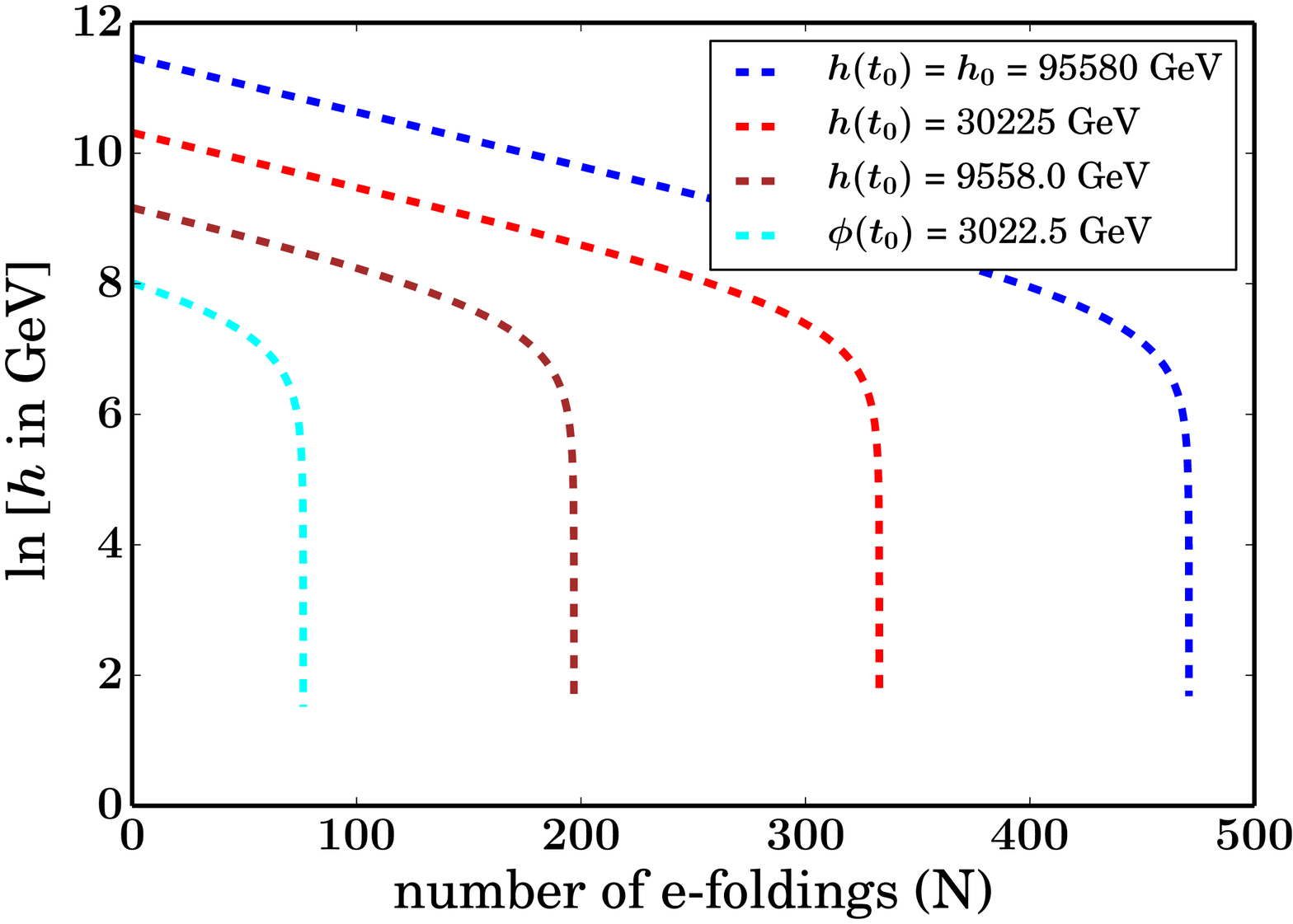}
\hfill
\includegraphics[width=.48\textwidth,origin=c,angle=0,natwidth=610,natheight=642]{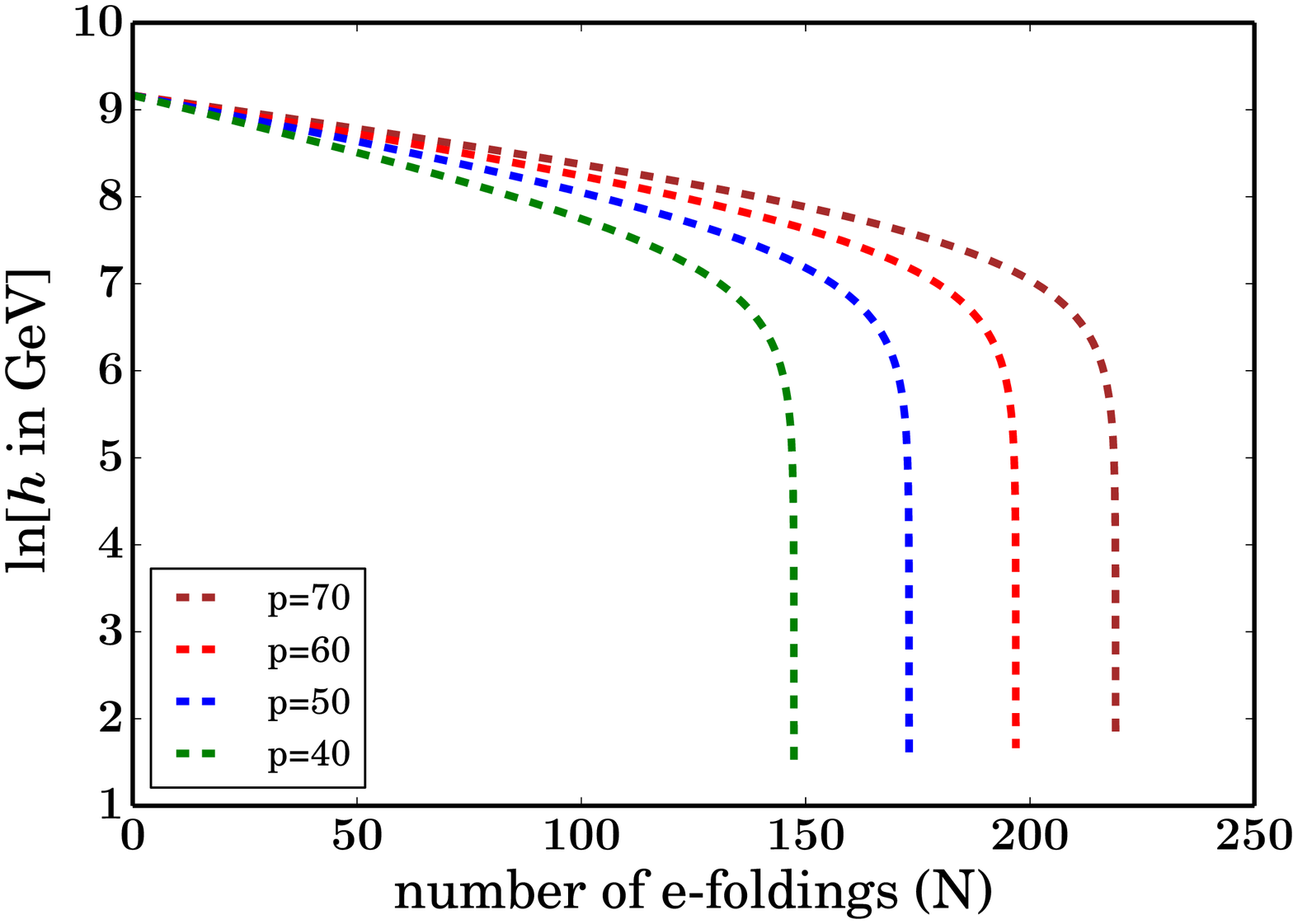}
% "\includegraphics" is very powerful; the graphicx package is already loaded
\caption{The evolution of Higgs field $h(t)$ (i) for different values of initial field value $h(t_0)$ (for p=60) and (ii) for different values of the parameter $\alpha_4(p)$.}
\label{fig:phivsN1} 
\end{figure}
%%%%%%%%%%%

%%%%%%%%%%%%%%%%%
\section{Power Spectrum}
\label{sec:Pspectrum}

In this section, we compute the scalar and tensor power-spectrum for
the Higgs inflation model, and compare it with the PLANCK data \cite{ade2014planck}. In
this section we use the results obtained by Hwang and Noh~\cite{hwang2000conserved}
for our analysis. At linear order, the scalar perturbations can be
simplified by writing in the uniform-field gauge. Mukhanov-Sasaki
equation in the Fourier domain is given by, see Ref. \cite{hwang2000conserved} for details:
\begin{equation}
\nu_k''+\left(c^2k^2-\frac{z''}{z}\right)\nu_k=0
\label{eq:fouriermode}
\end{equation}
where 
\begin{subequations}
\label{eq:MS-variables}
\begin{eqnarray}
z &=&a\sqrt{Q_{\mathcal{R}}} \\  
%%%%
Q_\mathcal{R}&=& \frac{\dot{\phi}^2+\ 12 H^2 \dot{f} \, \Gamma}{\left(H +\Gamma/2\right)^{2}} 
\qquad \quad \Gamma = \frac{8 H^2 \, \dot{f}}{1/\kappa+8H\dot{f}} \\
%%%%
c^2 &=& 
1- 256 \dot{f} \, \Gamma^2 \, \frac{\ddot{f}/\dot{f}-H-4 \dot{H}/\Gamma}{\dot{\phi}^2 + 12 H^2 \dot{f} \, \Gamma}
\end{eqnarray}
\end{subequations}
and prime denotes differentiation with respect to conformal time
$\eta$.  For the Higgs inflation, $c$ turns out to be approximately
constant with value slightly less than one and $Q_\mathcal{R}$ also is approximately a
constant, see the equations below,  for $p=60$:
\begin{equation}
c^2=\frac{1392 \MPlh^2 \left(60 \sqrt{10797} t_0 \sqrt{\frac{\MPlh^2}{\lambda  t_0^2}}+61 t \phi(t_0)^2-61 t_0\phi(t_0)^2\right)+3599 \sqrt{10797} {t_0} {\phi(t_0)}^2 
  \sqrt{\frac{\MPlh^2}{\lambda{t_0}^2}}}{1440 \MPlh^2
   \left(60 \sqrt{10797} {t_0} \sqrt{\frac{\MPlh^2}{\lambda  {t_0}^2}}+61 t \phi(t_0)^2-61 t_0 \phi(t_0)^2\right)+3599 \sqrt{10797} t_0\phi(t_0)^2 \sqrt{\frac{\MPlh^2}{\lambda {t_0}^2}}} 
   \label{cs}
\end{equation}
\begin{equation}
Q_\mathcal{R}=0.0017\left(\MPlh^2+\MPlh^2\left(\frac{1}{{0.235 \sqrt{\lambda } (t-{t_0})\MPlh} + \frac{24 \; \MPlh^2}{\phi(t_0)^2}}\right)\right) ~~\approx~~0.0017~\MPlh^2.
\label{Qs}
\end{equation}
\\
Hence for our model Eq.~(\ref{eq:fouriermode}) becomes :
\begin{equation}
\nu_k''+\left(c^2 k^2 - \frac{\sigma^2-1/4}{\eta^2}\right)\nu=0
\label{eq2:fouriermode}
\end{equation}
where 
\begin{equation}
 \sigma=\frac{3p-1}{2(p-1)}
 \end{equation}
 and the general solution for Eq.(\ref{eq2:fouriermode}) are a linear
 combination of Hankel functions.
\begin{equation}
\nu_k  = \sqrt{|\eta|}\left(AH_{\sigma}^{(1)}(c k|\eta|)+BH_{\sigma}^{(2)}(c k|\eta|)\right)
\end{equation}
Choosing the Bunch-Davies vacuum at past infinity ($c k |\eta|
\rightarrow \infty$), we get
\begin{equation}
 A=\sqrt{\frac{\pi}{4}}e^{i(2\sigma+1)\pi/4} \quad \mbox{and} \quad B=0 \, .
 \end{equation}
The power spectrum of scalar curvature perturbations
\begin{equation}
P_{\cal{R}}=\frac{k^3}{2 \pi^2}\left|\frac{\nu_k}{z}\right|^2
\end{equation}
evaluated when the modes leave the Hubble radius leads to 
%%%%%%%%%%%%%%%%%%%%%%%%%%%%%%%%%%%%%%%%%%%%%%
\begin{equation}
\label{eq:ScalarPowerSpectrum}
P_{\cal R}= {\cal C} \, k^{3 - 2\sigma}; \qquad 
{\cal C} = 2^{4(\sigma - 1)}c^{-2\sigma} \left(\frac{\Gamma(\sigma)}{\Gamma(3/2)}\right)^2\frac{1}{4 \pi^2} \left(\frac{a_0/t_0}{2\sigma-3}\right)^{2\sigma -1} \frac{1}{a_0^2 Q_{\mathcal{R}}} \, .
\end{equation}
The scalar spectral index $n_{\cal{R}} -  1= 3 - 2 \sigma$. 

The Fourier modes of the tensor perturbations satisfy, see Ref. \cite{hwang2000conserved} for details:
\begin{equation}
u''+\left(c_{\cal{T}}^2k^2-\frac{z_{\cal{T}}''}{z_{\cal{T}}}\right)u=0
\label{eq1:tensorfouriermode}
\end{equation}
where 
\begin{equation}
z_{\cal{T}}=a\sqrt{Q_g}; \quad  Q_g=\frac{1}{\kappa}+8H\dot{f},\quad c_{\cal{T}}^2 = \frac{1 +8 \kappa \ddot{f}}{1 + 8 \kappa H \, \dot{f}} \, .
\end{equation}
Like in the case of scalar-perturbations, $Q_g$ and $c_{\cal T}$ are
approximately constants during inflation.  So proceeding in the same
way as like scalar perturbations, the tensor power-spectrum is given
by:
\begin{equation}
\label{eq:TensorPowerSpectrum}
P_{\cal{T}}= {\cal C}_T k^{3-2\sigma_{\cal{T}}} \quad 
{\cal C}_T = 8\times 2^{4(\sigma_{\cal{T}} -1)}c_{\cal{T}}^{-2\sigma_{\cal{T}}} \left(\frac{\Gamma(\sigma_{\cal{T}})}{\Gamma(3/2)}\right)^2\frac{1}{4\pi^2} \left(\frac{a_0/t_0}{2 \sigma_{\cal{T}}-3}\right)^{2 \sigma_{\cal{T}} -1} 
\frac{1}{a_0^2 Q_{g}}
\end{equation}

The spectral index of scalar and tensor perturbations obey
$n_{\cal{R}}-1=n_{\cal{T}}$. The tensor to scalar ratio, which is
defined as
\begin{equation}
\label{eq:TensortoScalarratio}
r\equiv \frac{P_{\cal{T}}}{P_{\cal{R}}} =  8\times \left(\frac{c}{c_{\cal{T}}}\right)^{2\sigma} \frac{Q_{\cal{R}}}{Q_{g}}
\end{equation}
\vskip -0.8 cm

\subsection{Constraints from PLANCK}

PLANCK~\cite{ade2014planck} provides stringent constraints on the scalar
spectral index $n_\mathcal{R}=0.968 \pm 0.006$. Approximating inflation to be
generalized power-law (for large part during inflation),
Eq. (\ref{eq:ScalarPowerSpectrum}) leads to $p \approx 60$.  For $p
\approx 60$, from Eq.~(\ref{eq:TensortoScalarratio}), tensor to scalar
ratio turns out to be $r=0.012$. The above result is in agreement with
PLANCK constraint of $r<0.1$.  It is important to note that as like
other Higgs inflation model \cite{bezrukov2008standard}, our model predicts
that the contribution from the tensor is significantly smaller than
that of scalar perturbations.

Since our model does not have any free parameter, we can constraint
all the parameters of the model by comparing the model's predicted
power-spectrum and the observed power-spectrum values at the pivot
scale $k_*=0.05\mbox{Mpc}^{-1}$. The observed power-spectrum at the
pivot scale is $P_{\mathcal{R}}\left(k_*\right)=2.2\times10^{-9}$
\cite{ade2014planck}.  From Eq. (\ref{eq:ScalarPowerSpectrum}), the time
at which the perturbations exited the horizon radius during inflation
is given as
\begin{equation}
t_*=5.273\times 10^{-36}\; \mbox{secs}+ t_0 - 60 \sqrt{8 |f\left(h(t_0)\right)| \kappa} 
\label{eq3:t*}
\end{equation}
The field value at hubble crossing, $h_*\approx 10^{16}\;
\mbox{GeV}$. The above relations are the one of the main results of
our work, regarding which we would like to stress the following: (i)
Given that $\lambda_4$ is precisely
measured %\cite{Aad} i. e. $\lambda_4 = 0.0322$, using Eq. (\ref{eq:PLcoefficients})
leads to
\begin{equation}
h_0^2 t_0 \sim 5.7 \times 10^2 \MPlh \, .
\end{equation}
The above expression clearly shows that while $h_0^2 t_0$ can be
determined, however, $h_0$ and $t_0$ can not be determined
independently.  (ii) The time of exit of the perturbations depends on
the value of the Gauss-Bonnet parameter at the epoch of inflation. As
can be seen from above, $t_*$ depends on $t_0$ and $h_0$. (iii) The scale at the time perturbations left the Hubble scale is $H_*~ \approx 10^{12}\;\mbox{GeV}$.

\section{Discussions}
\label{sec:Conclusions}

In this work, we presented a model of inflation in which the Higgs
scalar is the inflaton. In our model, we have assumed Higgs is
non-minimally coupled to Gauss-Bonnet Gravity in 4-dimensions. We have
shown analytically that scalar field with $\phi^4$ potential term
leads to power-law inflation, however, adding a mass leads to the exit
of inflation. We have explicitly shown that the exit is close to the
Electroweak scale. Power-spectrum generated from our model is
consistent with the PLANCK observations.

There have been earlier attempts to look for an inflationary model at
electro-weak scales \cite{knox1993inflation,garcia1999nonequilibrium,krauss1999baryogenesis,german2001low,van2004electroweak,ross2016hybrid}.
Our model leads to inflation with exit at Electroweak scale due to
non-minimal coupling of the Higgs to Gauss-Bonnet gravity term. In the
model proposed by German et al Ref.~\cite{german2001low}, thermal effects lead to low-scale
inflation in supersymmetric or large extra dimensions.

It is intriguing that Gauss-Bonnet gravity which are higher-derivative
gravity corrections to Einstein gravity and hence are expected to have
strong effects only in the early universe.  However, we have
explicitly shown that Gauss-Bonnet gravity leads a dynamical model at
low-energies.

Our model is in spirit with Chaotic inflationary model of
Linde~\cite{linde1983chaotic}. Our model does not require an extension of
standard model but is a natural phenomenon within standard model at
the cost of a non minimal coupling of Higgs field with the
Gauss-Bonnet coupling.  Our analysis also indicate
  possible implications of the Gauss-Bonnet coupling at LHC.\footnote{Inflation reheating
  requires coupling of the inflaton with the standard model particles. More 
  precisely, to have efficient reheating the coupling has to be stronger so
  that the energy gets transferred from inflaton to standard model particles. 
  If the exit happens close to EW scale, then this implies that the reheating 
  process can provide some hint about the coupling of the inflaton with 
  standard model.}

As mentioned in the previous section, PLANCK observations constrain
all parameters of our model, however, it does not constrain the epoch
of inflation. Using the non-Gaussianity constraints of PLANCK may help
to break the degeneracy between $h_0$ and $t_0$.  This is currently
under investigation.
%%%%%%%%%%%%%%%%%%%%%%%%%%%%%%%%  
\section{Acknowledgements}
%%%%%%%%%%%%%%%%%%%%%%%%%%%%%%%%
The work is supported by Max Planck-India Partner Group on Gravity and
Cosmology.  JM is supported by UGC Fellowship. JM thanks Krishnamohan
Parattu for useful discussions.

\section{Appendix: Numerical Evaluation}
In order to solve the time evolution of the Higgs field $h(t)$, we
obtain the expression for $\frac{d^2h}{dt^2}$ as a function only of
$h$ and $\frac{dh}{dt}$.  We then use RK4 to numerically evaluate
$h(t)$.

We solve the Hubble constant H from (\ref{energyconstraint}). The
cubic equation leads to three solutions for H, given by:
\begin{subequations}
\begin{eqnarray}
H1&=&\frac{B^{\frac{1}{3}}}{24 k \frac{df(t)}{dt}}+\frac{1}{24 k\frac{df(t)}{dt} B^{\frac{1}{3}}}-\frac{1}{24 k \frac{df(t)}{dt}}\label{H1}\\
H2&=&\frac{-B^{\frac{1}{3}}}{48 k \frac{df(t)}{dt}} - \frac{1}{48 k\frac{df(t)}{dt} B^{\frac{1}{3}}}-\frac{1}{24 k \frac{df(t)}{dt}}
+I\sqrt{3}\left(\frac{B^{\frac{1}{3}}}{48 k \frac{df(t)}{dt}} - \frac{1}{48 k\frac{df(t)}{dt} B^{\frac{1}{3}}}\right)\label{H2}\\
H3&=&\frac{-B^{\frac{1}{3}}}{48 k \frac{df(t)}{dt}} - \frac{1}{48 k\frac{df(t)}{dt} B^{\frac{1}{3}}}-\frac{1}{24 k \frac{df(t)}{dt}}
-I\sqrt{3}\left(\frac{B^{\frac{1}{3}}}{48 k \frac{df(t)}{dt}} - \frac{1}{48 k\frac{df(t)}{dt} B^{\frac{1}{3}}}\right)\label{H3}
\end{eqnarray}
\end{subequations}

where
\begin{eqnarray*}
B &=& 144\, \left( {\frac {d}{dt}}f \left( t \right)  \right) ^{2}{k}^{3}
 \left( {\frac {d}{dt}}h(t)  \right) ^{2}+288\,
 \left( {\frac {d}{dt}}f \left( t \right)  \right) ^{2}{k}^{3}V
 \left( t \right) -1+\\
 &\;& 12\,\sqrt {2} \left( {\frac {d}{dt}}f \left( t
 \right)  \right) {k}^{2}\sqrt {A}\\
A &=& \frac {1}{k}\left(72\, \left( {\frac {d}{dt}}f \left( t \right)  \right) ^{2} {k}^{3} \left( {\frac {d}{dt}}h(t)  \right) ^{4}+288\,
 \left( {\frac {d}{dt}}f \left( t \right)  \right) ^{2}{k}^{3} \left( 
{\frac {d}{dt}}h(t)  \right) ^{2}V \left( t \right) -\right.\\
 &\;&\left.\left( {\frac {d}{dt}}h(t)  \right) ^{2}+
 288\, \left( {\frac {d}{dt}}f \left( t \right)  \right) ^{2}{k}^{3} \left( 
V \left( t \right)  \right) ^{2}-2\,V \left( t \right) \right)
\end{eqnarray*}

Among the three solutions for H, {\sl only} one solution satisfy the
eqns. (\ref{eq:threeequations}) all through the evolution and that is
the physical solution.  Substituting the corresponding solution in
(\ref{eom}), we get the second order differential equation in
$h(t)$. From the fact that we can split $\frac{dH}{dt}$ as a sum of
$H_{t1}$ and $H_{t2} \frac{d^2 h}{dt^2}$, where $H_{t1}$ and $H_{t2}$
are independent of $\frac{d^2h}{dt^2}$. we have:
\begin{equation}
\frac{d^2 h}{dt^2}=\frac{24 H^2\frac{df(h)}{dt} H_{t1}+24H^4\frac{df(h)}{dt}-\frac{dV(h)}{dt}-3\left(\frac{dh}{dt}\right)^2 H}{\frac{dh}{dt}-24 H^2\frac{df(h)}{dt}H_{t2}}
\end{equation}
\linebreak
\centering{\noindent\rule{15cm}{0.4pt}}

 \bibliographystyle{elsarticle-num} 
\bibliography{bibdata}{}

\end{document}